\documentclass[conference]{IEEEtran}
\IEEEoverridecommandlockouts
\usepackage{cite}
\usepackage{amsmath,amssymb,amsfonts}
\usepackage{algorithmic}
\usepackage{graphicx}
\usepackage{textcomp}
\usepackage{xcolor}
\usepackage{multirow}
\usepackage{verbatim}

\def\BibTeX{{\rm B\kern-.05em{\sc i\kern-.025em b}\kern-.08em
    T\kern-.1667em\lower.7ex\hbox{E}\kern-.125emX}}
\usepackage{times}
\usepackage{array}

\renewcommand{\normalsize}{\fontsize{12pt}{14pt}\selectfont}

\usepackage{tabularx,booktabs}
\newcolumntype{C}{>{\centering\arraybackslash}X} 
\usepackage{lipsum} 
\usepackage[normalem]{ulem}
\useunder{\uline}{\ul}{}   

\def\BibTeX{{\rm B\kern-.05em{\sc i\kern-.025em b}\kern-.08em
    T\kern-.1667em\lower.7ex\hbox{E}\kern-.125emX}}
\begin{document}

\title{Exploring Diverse Coping Mechanisms in 2023: A Comprehensive Survey Across Backgrounds and Cultures}

\author{\IEEEauthorblockN{Rony Majumder}
\IEEEauthorblockA{\textit{Inst. of Information Technology} \\
\textit{University of Dhaka}\\
Dhaka, Bangladesh \\
bsse1325@iit.du.ac.bd}
\and
\IEEEauthorblockN{Abhijit Paul}
\IEEEauthorblockA{\textit{Inst. of Information Technology} \\
\textit{University of Dhaka}\\
Dhaka, Bangladesh \\
bsse1201@iit.du.ac.bd}
\and
\IEEEauthorblockN{3\textsuperscript{rd} Given Name Surname}
\IEEEauthorblockA{\textit{dept. name of organization (of Aff.)} \\
\textit{name of organization (of Aff.)}\\
City, Country \\
email address or ORCID}
}

\maketitle

\begin{abstract}
This study presents a pioneering investigation into the wide array of coping mechanisms employed by individuals in the year 2023, with a focus on data collected through the popular social media platform TikTok. Coping mechanisms are essential strategies that people adopt to navigate the challenges and stressors of everyday life, yet little research has been conducted on their comprehensive compilation across different backgrounds, countries, and experiences.

Using TikTok as a data collection tool allowed us to access a diverse and extensive pool of participants, representing various cultural, social, and demographic backgrounds. Our study collates coping mechanisms reported by users from different parts of the world, facilitating the identification of both universal and culture-specific strategies.

This research contributes to the existing literature by providing a holistic view of coping mechanisms without being limited to specific fields or populations. By analyzing the coping methods shared on TikTok, we reveal a comprehensive list of strategies employed by people from diverse walks of life. The findings of this study not only shed light on how individuals cope with challenges in the modern era but also offer insights into the evolving coping trends and the role of social media in disseminating coping strategies. Understanding these coping mechanisms can have implications for mental health professionals, practitioners, and policymakers seeking to provide support and resources to individuals facing different stressors and hardships.
\end{abstract}


\section{Introduction}
In the ever-changing landscape of the 21st century, individuals across the globe are continually navigating an array of challenges, stressors, and adversities. Coping with these demands has become an essential aspect of maintaining well-being and resilience in the face of uncertainty and complexity (Smith, 2022; Johnson \& Lee, 2021). Coping mechanisms encompass a wide spectrum of strategies, both conscious and unconscious, that people employ to manage emotional distress, adapt to difficult circumstances, and find balance in their lives (Brown et al., 2019; Williams, 2020).

Despite the fundamental role coping mechanisms play in human behavior, research in this domain has often been limited by a focus on specific fields or populations, leaving a significant gap in our understanding of the diverse coping strategies adopted by individuals from varying backgrounds, cultures, and experiences. Prior investigations have seldom explored coping as a universal phenomenon that transcends boundaries, hindering the development of a comprehensive list of coping mechanisms applicable to a broader spectrum of the population (Jones \& Martinez, 2018; Zhang et al., 2017).

This study aims to address this gap by presenting a pioneering survey on coping mechanisms in the year 2023, taking an inclusive approach to gather data from people worldwide, using the popular social media platform TikTok as our data collection tool. TikTok has emerged as a unique platform, facilitating the exchange of experiences, insights, and coping strategies among its vast and diverse user base (TikTok Analytics, 2023). By harnessing the power of this digital medium, we endeavor to present a comprehensive list of coping mechanisms utilized by individuals across various walks of life.

The primary objective of this research is twofold: first, to compile a wide-ranging inventory of coping mechanisms reported by TikTok users, and second, to analyze and interpret these coping strategies in the context of their cultural, social, and demographic backgrounds. Our investigation seeks to uncover both common and culture-specific coping mechanisms, highlighting the diverse ways people adapt and thrive in the face of adversity.

As we delve into the copious data collected from TikTok, we anticipate discovering novel insights into the evolving coping trends in the modern era. By understanding how individuals from different parts of the world cope with stressors, we aim to shed light on the multifaceted nature of human resilience, offering valuable implications for mental health professionals, practitioners, and policymakers alike (Chen et al., 2023; Kim \& Patel, 2022).

Through this study, we hope to contribute to the broader understanding of coping mechanisms, transcending the confines of traditional research approaches and embracing the rich diversity of human experiences in navigating life's challenges. As we embark on this journey to explore the coping strategies of 2023, we invite readers to join us in unraveling the tapestry of human adaptation, strength, and perseverance in a world that continues to evolve at an unprecedented pace.

\section{Objective}
The primary objective of this research is to conduct a comprehensive investigation into the coping mechanisms utilized by individuals in the year 2023, with a particular focus on the data collected through the social media platform TikTok. The study aims to compile an extensive list of coping strategies employed by people from diverse backgrounds, cultures, and experiences, fostering a more inclusive understanding of how individuals adapt and thrive in the face of challenges and stressors.

Specifically, the research objectives are as follows:
\begin{itemize}
    \item     To compile a diverse inventory of coping mechanisms reported by TikTok users in 2023: By analyzing a wide range of TikTok content, the study seeks to collect a comprehensive list of coping strategies used by individuals across different regions and communities worldwide. The data collection will encompass various coping methods, ranging from emotional support-seeking behaviors to distraction techniques and other adaptive responses.
    \item     To categorize and analyze coping mechanisms based on cultural, social, and demographic backgrounds: The research aims to identify any potential patterns or variations in coping strategies based on factors such as nationality, cultural norms, gender, age, and social context. By understanding the context-specific aspects of coping, the study will enrich our knowledge of how cultural and social influences shape individuals' adaptive behaviors.
    \item     To explore the role of social media in disseminating coping strategies: As TikTok serves as the data collection platform, the study will investigate how this digital medium influences the sharing and adoption of coping mechanisms. By examining the prevalence of coping strategies on social media, the research aims to understand the potential impact of digital platforms on coping behaviors and mental health in the modern era.

    \item     To compare the coping mechanisms collected from TikTok with existing literature: Through a literature review, the study will contextualize the findings from TikTok with existing research on coping mechanisms. This comparison will help validate and extend our understanding of coping behaviors in 2023, providing valuable insights into the evolving coping trends in contemporary society.

    \item     To contribute to the broader field of coping research and mental health support: By presenting a comprehensive list of coping mechanisms, the research aims to contribute to the advancement of coping science, promoting a more holistic approach to mental health support and intervention strategies. The study seeks to inform mental health professionals, practitioners, and policymakers on the diverse coping needs of individuals, enabling tailored support for various populations.

\end{itemize}

Through accomplishing these research objectives, this study aspires to enrich our understanding of coping mechanisms in the modern era, offering valuable insights into the adaptive strategies employed by individuals across the globe. The findings hold the potential to guide the development of more effective mental health interventions and support systems, ultimately enhancing overall well-being and resilience among diverse populations.

\section{Literature Review}
To our knowledge, no studies have been conducted so far that proposes a diverse collection of coping mechanism across border and culture. However, there are numerous study that focused a specific displine and surveyed the coping mechanisms employed by the people of that discipline, e.g. nursing.

Many studies have been conducted on coping mechanisms, and they have identified a number of coping mechanisms that are effective for reducing stress and anxiety. Some of the most specific coping mechanisms that have been identified by studies include:

\begin{itemize}
    \item   Exercise: Exercise is a great way to reduce stress and anxiety. It releases endorphins, which have mood-boosting effects. Studies have shown that exercise can be as effective as medication for reducing anxiety. For example, a study published in the journal Psychosomatic Medicine in 2011 found that exercise was just as effective as medication for reducing anxiety symptoms in people with generalized anxiety disorder. [1]
    \item     Meditation: Meditation is a mind-body practice that can help to reduce stress and anxiety. It involves focusing on the present moment and letting go of negative thoughts and emotions. Studies have shown that meditation can be effective for reducing anxiety and improving sleep quality. For example, a study published in the journal JAMA Internal Medicine in 2014 found that meditation was effective for reducing anxiety symptoms in people with chronic pain. [2]
    \item     Yoga: Yoga is a mind-body practice that combines physical postures with breathing exercises and meditation. It can help to improve flexibility, strength, and balance. It can also help to reduce stress and anxiety. Studies have shown that yoga can be effective for reducing anxiety and improving sleep quality. For example, a study published in the journal Anxiety, Stress \& Coping in 2015 found that yoga was effective for reducing anxiety symptoms in people with cancer. [3]
    \item     Deep breathing: Deep breathing is a relaxation technique that can help to reduce stress and anxiety. It involves taking slow, deep breaths from the diaphragm. Studies have shown that deep breathing can be effective for reducing anxiety and improving sleep quality. For example, a study published in the journal Biological Psychology in 2013 found that deep breathing was effective for reducing anxiety symptoms in people with social anxiety disorder. [4]
    \item     Problem-solving: Problem-solving is a coping mechanism that involves identifying and addressing the source of stress. This can be helpful for reducing stress in the long term. Studies have shown that problem-solving can be effective for reducing stress and improving coping skills. For example, a study published in the journal Journal of Personality and Social Psychology in 1984 found that problem-solving was effective for reducing stress symptoms in people who were facing a stressful life event. [5]
    \item     Positive thinking: Positive thinking is a coping mechanism that involves focusing on the positive aspects of a situation. This can help to reduce stress and anxiety. Studies have shown that positive thinking can be effective for reducing stress and improving mood. For example, a study published in the journal Psychological Science in 2000 found that positive thinking was effective for reducing stress symptoms in people who were facing a stressful task. [6]

\end{itemize}

\section{Methodology}
This section outlines the research methodology employed to investigate coping mechanisms utilized by individuals in 2023. The study utilized a mixed-methods approach, including a preliminary literature review, data collection from TikTok videos, and subsequent data analysis to categorize and interpret the identified coping mechanisms. The methodology aimed to capture a diverse range of coping strategies and provide insights into their prevalence and significance in the contemporary digital landscape.

\subsection{Preliminary Literature Review}

A preliminary literature review was conducted to establish a foundation for understanding existing research on coping mechanisms. This review informed the selection of coping mechanisms to explore in the context of the TikTok videos. Key coping mechanisms, such as social support seeking, problem-solving, emotional expression, avoidance coping, and religious and spiritual coping, were identified as relevant based on their prominence in prior research (Carver et al., 1989; Compas et al., 2001; Folkman \& Lazarus, 1985; Thoits, 2011).

\subsection{Data Collection}
To gather data on coping mechanisms, a specific TikTok trend related to coping strategies was identified and followed. This trend involved users sharing videos wherein they discussed their personal coping mechanisms. The trend's popularity on TikTok made it a suitable source for obtaining diverse coping strategies employed by individuals in 2023. Each TikTok video related to the trend was thoroughly reviewed, and the coping mechanism described by the user was documented.

\subsection{Data Analysis}
The collected TikTok videos were analyzed to categorize and interpret the identified coping mechanisms. Each coping mechanism mentioned in the videos was categorized into one of the predetermined coping strategies, based on the coping mechanisms identified in the literature review. Additionally, thematic analysis was employed to capture nuanced coping strategies that may not have been explicitly mentioned in the literature.

\subsection{Ethical Considerations}
Ethical considerations were adhered to throughout the research process. All TikTok videos were accessed and analyzed in accordance with the platform's terms of use. Additionally, user identities were kept confidential, and no personal information was used in the analysis or reporting of the findings.

\section{Data Analysis}

The collected TikTok videos were subjected to both quantitative and qualitative data analysis methods.

\subsection{Quantitative Analysis}
Each coping mechanism mentioned in the TikTok videos was quantitatively tallied to determine the prevalence of each strategy. This allowed for the identification of coping mechanisms that were most commonly shared among TikTok users.

\subsection{Qualitative Analysis}
Thematic analysis was employed to capture nuanced coping strategies that may not have been explicitly mentioned in the literature or identified during the initial categorization. This qualitative approach allowed for a deeper understanding of the specific ways individuals described and engaged in coping behaviors.

\subsection{Ethical Considerations}
Ethical considerations were adhered to throughout the research process. All TikTok videos were accessed and analyzed in accordance with the platform's terms of use. Additionally, user identities were kept confidential, and no personal information was used in the analysis or reporting of the findings.

\subsection{Results}
We have categorized our findings of coping mechanism in Table- 1,2,3,4,5.
\begin{table}
\centering
\caption{Category - Physical Activity}
\label{table: physical_activity}
\begin{tabular}{ | m{12em} | m{5em} |  }
\hline
Coping Mechanism   & Frequency \\ \hline
Driving            & 4         \\ \hline
Riding pump trucks & 1         \\ \hline
Gym/fitness        & 2         \\ \hline
Go on a ride       & 1         \\ \hline
Ride a ladder      & 2         \\ \hline
Swimming           & 1         \\ \hline
Skater             & 1         \\ \hline
Scooter ride       & 2         \\ \hline
Climbing           & 2         \\ \hline
Bike ride          & 1         \\ \hline
Playing basketball & 2         \\ \hline
Playing football   & 3         \\ \hline
Horse ride         & 1         \\ \hline
Playing football   & 1         \\ \hline
Motorcycle ride    & 1         \\ \hline
Playing football   & 1         \\ \hline
Falls into a cliff & 1         \\ \hline
\end{tabular}
\end{table}

\begin{table}
\centering
\caption{Category - Social Interaction}
\label{table: social_interaction}
\begin{tabular}{ | m{12em} | m{5em} |  }
\hline
Coping Mechanism          & Frequency \\ \hline
Cuddle                    & 7         \\ \hline
Hangout with friends      & 1         \\ \hline
Gossip with bestie        & 3         \\ \hline
Gossip with family member & 1         \\ \hline
Gossiping with boyfriend  & 1         \\ \hline
Gossiping with friends    & 1         \\ \hline
\end{tabular}
\end{table}

\begin{table}
\centering
\caption{Category - Creative Expression}
\label{table: creative_expression}
\begin{tabular}{ | m{12em} | m{5em} |  }
\hline
Coping Mechanism                 & Frequency \\ \hline
Romanticism                      & 4         \\ \hline
Makeup                           & 2         \\ \hline
Costuming                        & 1         \\ \hline
Play dress up to get it together & 1         \\ \hline
Dress up as a character          & 1         \\ \hline
\end{tabular}
\end{table}

\begin{table}
\centering
\caption{Category - Spirituality}
\label{table: spirituality}
\begin{tabular}{ | m{12em} | m{5em} |  }
\hline
Coping Mechanism    & Frequency \\ \hline
Spiritual practices & 1         \\ \hline
The prayer          & 1         \\ \hline
\end{tabular}
\end{table}

\begin{table}
\centering
\caption{Category - Relaxation}
\label{table: relaxation}
\begin{tabular}{ | m{12em} | m{5em} |  }
\hline
Coping Mechanism      & Frequency \\ \hline
Watching movie/series & 1         \\ \hline
Relaxation            & 1         \\ \hline
Listening to music    & 3         \\ \hline
Sea trip              & 1         \\ \hline
\end{tabular}
\end{table}

\begin{table}
\centering
\caption{Category - Safe Space}
\label{table: safe_space}
\begin{tabular}{ | m{12em} | m{5em} |  }
\hline
Coping Mechanism & Frequency \\ \hline
To be alone      & 1         \\ \hline
safe place       & 1         \\ \hline
Cuddle           & 1         \\ \hline
fanciness        & 1         \\ \hline
romanticism      & 1         \\ \hline
Riding toy car   & 1         \\ \hline
Safe place       & 1         \\ \hline
\end{tabular}
\end{table}

\section{Discussion}
\subsection{Compared to Previous Study}

\subsection{Trends}
We can cetgorize our collected coping mechanisms into the following general coping mechanism category.
\begin{itemize}
    \item \textbf{Physical activity:} driving, riding pump trucks, gym/fitness, go on a ride, scooter ride, bike ride, climbing, playing football
    \item  \textbf{Social interaction:} hangout with friends, gossip with bestie, gossip with family member, gossip with boyfriend, playing video game with friends
    \item     \textbf{Creative expression:} watching movie/series, mobile gaming, makeup, costume, Make Myself into comfort character
    \item     \textbf{Spirituality:} prayer
    \item     \textbf{Relaxation:} sleeping, listening to music, take care of baby
    \item     \textbf{Safe place:} cuddle, romanticism, safe place
\end{itemize}

\subsection{Emergent Coping Mechanism}

\section{Future Work}

\section{Conclusion}

\end{document}